\documentclass[aps,prl,reprint,superscriptaddress,twocolumn]{revtex4-1}

\usepackage{graphicx}
\usepackage{dcolumn}
\usepackage{bm}
\usepackage{amsmath}

\usepackage{color}

\begin{document}


\title{%
Spin Dependent Collision of Ultracold Metastable Atoms%
}

\author{Satoshi Uetake}%
 \thanks{Present address: Department of Physics, Okayama University,
 3-1-1, Tsushima-naka, Kita-ku, Okayama 700-8530, Japan}
 \email{uetake@scphys.kyoto-u.ac.jp}
 \affiliation{Department of Physics, Graduate School of Science,
 Kyoto University, Kyoto 606-8502, Japan}
 \affiliation{CREST, Japan Science and Technology Agency, Chiyoda,
 Tokyo 102-0075, Japan} 
\author{Ryo Murakami}%
 \affiliation{Department of Physics, Graduate School of Science,
 Kyoto University, Kyoto 606-8502, Japan}
\author{John M. Doyle}%
 \affiliation{Department of Physics, Harvard University, Cambridge,
 Massachusetts 02138, USA} 
\author{Yoshiro Takahashi}%
 \affiliation{Department of Physics, Graduate School of Science,
 Kyoto University, Kyoto 606-8502, Japan}
 \affiliation{CREST, Japan Science and Technology Agency, Chiyoda,
 Tokyo 102-0075, Japan} 

\date{\today}

\begin{abstract}
 Spin-polarized metastable atoms of ultracold ytterbium are trapped at
 high density and their inelastic collisional properties are measured. We
 reveal that in collisions of Yb($^3P_2$) with Yb($^1S_0$) there is
 relatively weak inelastic loss, but with a significant
 spin-dependence consistent with Zeeman sublevel changes as being the
 dominant decay process. This is in strong contrast to our
 observations of Yb($^3P_2$)--Yb($^3P_2$) collisional loss, which are,
 at low field, much more rapid and have essentially no spin
 dependence. 
 Our results 
 give a guideline to use the $^3P_2$
 states in many possible applications.
\end{abstract}


\maketitle

Atoms with alkaline-earth-metal-like electronic structure  
are under extensive study, partly due to their promise for
a number of key applications.
For example, the 
use of long-lived metastable $P$ states (as well as the ground $S$
state) has been explored as a useful quantum computing 
platform \cite{KShibata:ApplPhysB2009,%
AJDaley:PRL2008,%
RStock:PRA2008,%
ADerevianko:PRA2004,%
AVGorshkov:PRL2009}.
Also, the ultranarrow $^1S_0$--$^3P_0$ atomic resonance in a ``magic
wavelength'' optical lattice may be highly competitive as a new
optical frequency standard \cite{HKatori:PRL2003}.
Finally, in the area of quantum simulation, there are several
theoretical studies of the use of $^3P_J$ ($J=0,2$) atoms for studies
of Hamiltonians with both spin and orbital degrees of 
freedom \cite{AVGorshkov:NaturePhys2010,CXu:PRB2010},
implementation of Abelian artificial gauge potentials
\cite{FGerbier:NJourPhys2010}, or 
simulation of Kondo lattice model \cite{MFoss-Feig:PRA2010}.

Collisions of metastable $P$ state atoms are interesting not only from a
fundamental interactions viewpoint, and as the determining factor in
collisional cooling schemes, but also as a crucial mechanism in several key
applications. Quantum gate phase imprinting via collisions has been 
proposed \cite{DJaksch:PRL1999,AVGorshkov:PRL2009,AJDaley:PRL2008}, 
as
well as the exchange interaction in Kondo lattice model simulations
\cite{MFoss-Feig:PRA2010}. 
Collisions can not only enable new physics but also inhibit
desired applications. For example, in optical frequency standards
collisional shifts can limit their accuracy, as well as lead to inelastic
losses that destroy the atomic sample, possibly on a time scale shorter than
the desired interrogation time. Recent studies with metastable
alkaline-earth-metal-like structure atoms showed inelastic collision rates
as high as $10^{-17}$ to $10^{-16}$~m$^3/$s in $^3P_J$--$^3P_J$ collisions
in Yb \cite{AYamaguchi:PRL2008%
} and Sr \cite{ATraverso:PRA2009,%
ChLisdat:PRL2009%
} atoms. These rates, almost as high as the estimated
elastic collisional rates, make the use of $^3P_J$ atoms difficult.

The role of the collisions between the ground ($^1S_0$) and excited
($^3P_J$) atoms is crucially important in many approaches
\cite{AJDaley:PRL2008,AVGorshkov:PRL2009,MFoss-Feig:PRA2010}.
In this respect, it is important to investigate the  properties of
collisions between the metastable triplet states and the ground state
\cite{ChLisdat:PRL2009,NDLemke:PRL2011}.  
In addition, the study of anisotropically interacting cold collisions is
now a broadening area of study and includes the rare-earth atoms
and polar molecules
\cite{%
CIHancox:PRL2005,RVKrems:PRL2005,MJLu:PRA2009,
MLu:PRL2011,MLu:PRL2012,SKotochigova:PhysChemChemPhys2011,%
APetrov:arXiv2012,KAikawa:PRL2012,
LPParazzoli:PRL2011
}.

In this Letter, we report the measurement of the inelastic rate
constants for Yb in both the $^3P_2$--$^3P_2$ and $^1S_0$--$^3P_2$
systems at several magnetic fields below 1 gauss. 
In particular, we observe strong spin dependence in the inelastic
rates for Yb($^3P_2$) in the 
$^1S_0$--$^3P_2$ collisional system, with higher energy Zeeman
sublevels having higher inelastic rates. This strongly suggests that
the inelastic loss is dominated by Zeeman sublevel changing processes
(``$m$-changing collisions''). These may 
be induced by the Landau-Zener transition between the entrance s-wave
channel and the higher partial waves with lower magnetic sublevels
\cite{VKokoouline:PRL2003,RSantra:PRA2003}.
In contrast, in the $^3P_2$--$^3P_2$ collisions we observe a much
higher inelastic rate that is spin-independent. This is consistent
with fine structure changing processes (``$J$-changing collisions'')
or 
principal quantum number changing (``PQNC'') processes. 
Our results represent a new, detailed study of $^3P_2$ physics,
and also 
provide a roadmap for the use of the important $^3P_2$ state and
challenge theory to provide a quantitative explanation of this
anisotropic collision physics.  

\begin{figure}
\includegraphics{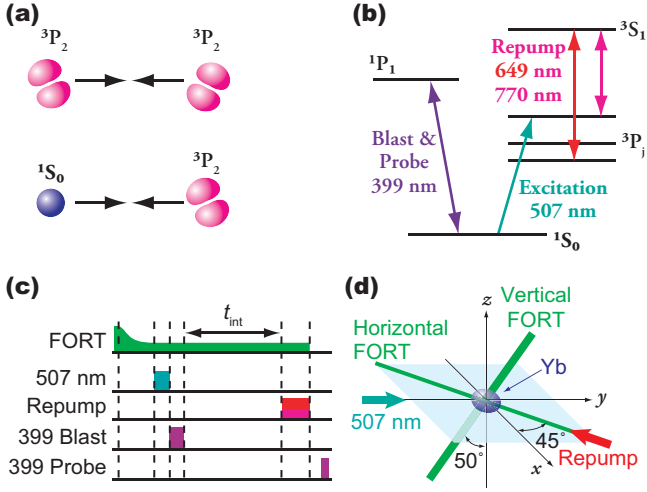}
\caption{(color online) (a) Schematics of the collision
 measurement. We studied both $^3P_2$--$^3P_2$ and
 $^1S_0$--$^3P_2$ collisional properties. (b) Energy level
 diagram of ytterbium. 
 (c) Timing diagram for measurement of collision dynamics of
 $^3P_2$--$^3P_2$ and $^1S_0$--$^3P_2$ collision system. 
 For measurement of $^3P_2$--$^3P_2$ collision system, unexcited $^1S_0$
 atoms are immediately removed from the trap by applying a strong
 399~nm laser. 
 For measurement of $^1S_0$--$^3P_2$ collision, the blast laser is
 irradiated after interaction time.
 (d) Schematic diagram of experimental setup. 
 Directions of 399 blast and probe lasers (not shown) are same as that
 of 507~nm excitation laser. 
}
\label{fig:timing}
\end{figure}

In a previous experiment \cite{AYamaguchi:PRL2008}, we studied $^3P_2$
collisions via indirect excitation of the $^3D_2$ state to
$^3P_2$. This led to a spin-unpolarized sample at the relatively high
temperature of around 40~$\mu$K. Here we prepare spin polarized samples at
a much lower temperature, below 1~$\mu$K, as described below.  
The experimental procedure is as follows. 
Yb atoms in a thermal beam generated from an oven at 375$^\circ$C are
decelerated by a Zeeman slower with a strong $^1S_0$--$^1P_1$
transition at 399~nm, and then are loaded into the intercombination
($^1S_0$--$^3P_1$) MOT at
556~nm \cite{TKuwamoto:PRA1999,SUetake:ApplPhysB2008}.  
Figure \ref{fig:timing}(b)--(d) show corresponding energy level
diagram of Yb, timing diagram of the experiment, and
schematic diagram of experimental setup, respectively.
The atoms are transferred from the MOT to a crossed FORT at 532~nm and 
evaporatively cooled.
The number of atoms is typically $3\times 10^5$.
The BEC
transition temperature ($T_c$) of our trap is  400~nK. All the work reported in
this paper 
is done 
at 480~nK, just above $T_c$, in order to maintain a
stable atom number. 
Our spin-selective 
excitation employs the very
narrow linewidth $^1S_0$--$^3P_2$ transition at 507 nm
\cite{AYamaguchi:ApplPhysB2008,AYamaguchi:NewJPhys2010}. 
The
excitation efficiency to $^3P_2$ is about 10\%, typically leaving about
$2\times 10^4$ $^3P_2$ atoms in the trap.  
During the excitation we apply a small bias magnetic field
$B_{\text{bias}}$ to spectroscopically split magnetic sublevels $m_J$
of $^3P_2$ state. 
The applied magnetic field varies from 215~mG to 848~mG, corresponding
to the Zeeman splitting of 0.45~MHz to 1.8~MHz. 
The excitation laser linewidth is measured to be less than 1~kHz and
Doppler broadening is about 20~kHz. This is much less than the
Zeeman splitting of 450~kHz at our lowest magnetic field of 215~mG, thus
ensuring selective creation of spin-polarized samples of atoms in a single
Zeeman sublevel.

We perform study of two different collisional systems,
$^3P_2$--$^3P_2$ and 
$^1S_0$--$^3P_2$. 
For the former, 
the leftover $^1S_0$ atoms are immediately removed by strong excitation at
399~nm (i.e. blast laser) [Fig.~\ref{fig:timing}(c)]. 
After interaction time $t_{\text{int}}$ (where the $^3P_2$ atoms
collide with other atoms and undergoes inelastic loss), $^3P_2$ atoms
are repumped back to the $^1S_0$ state by 770~nm and 649~nm repumping
lasers which are resonant to the $^3P_2$--$^3S_1$ and $^3P_0$--$^3S_1$
transitions. A few milliseconds is required for complete repumping.  
Finally the number of repumped ($^1S_0$) atoms are measured by
absorption imaging using the $^1S_0$--$^1P_1$ transition.  
For the latter (i.e. $^1S_0$--$^3P_2$ collision measurement), 
we simply leave the leftover $^1S_0$ atoms in the trap.
Then the blast laser is irradiated after interaction time: just before
the repumping. 

\begin{figure*}
\includegraphics[width=0.98\textwidth]{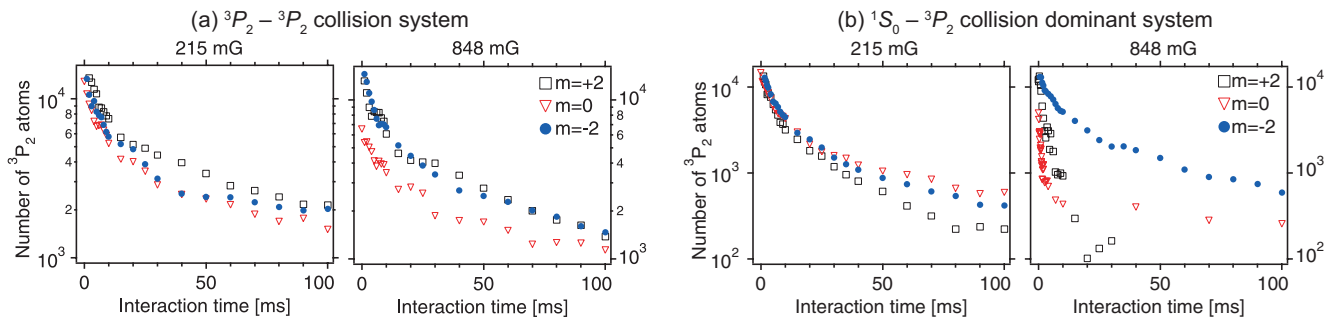}
 \caption{(color online) Observed decay of trapped $^3P_2$ atoms as a
 function of time. (a) Time evolution of the number of $^3P_2$
 atoms without $^1S_0$ atoms. Decay rates of different magnetic
 sublevels $m_J$ at different magnetic field strengths are essentially
 the same. (b) Time evolution of the number of $^3P_2$ atoms with
 $^1S_0$ atoms. In this case the number of $^1S_0$ atoms are 10 times
 larger than that of $^3P_2$ atoms. 
 Thus the $^3P_2$ atoms dominantly collides with $^1S_0$ atoms.
 The behavior is markedly different
 from that of $^3P_2$--$^3P_2$ collision system. Decay rates are
 strongly spin-dependent. Note that we measured decay for all
 magnetic components in each bias magnetic field of 215, 307, 407,
 596, and 848~mG. Part of data is shown in these graphs.}
\label{fig:decay}
\end{figure*}

Inelastic atom-atom collisions in various channels are clearly
observed for all Zeeman states.  
A selection of these state-dependent decay curves (those at highest
and lowest magnetic field) are presented in Fig.~\ref{fig:decay}. 
Figure \ref{fig:decay}a (b) shows the time evolution of the number of
$^3P_2$ atoms without (with) the $^1S_0$ atoms.
Thus the decay of atoms in Fig.~\ref{fig:decay}a is due only to
$^3P_2$--$^3P_2$ collisions. The measurements reveal a very high
inelastic loss rate (approximately the same as the estimated elastic
rate) that is essentially independent of both $m_J$ and magnetic field
strength. This behavior is consistent with the scenario that the
dominant decay process is $J$-changing, which was inferred by our
previous measurement \cite{AYamaguchi:PRL2008}. It may also be
due to PQNC collisions, as described in 
Ref. \cite{ATraverso:PRA2009}.

\begin{figure*}
\includegraphics[width=0.99\textwidth]{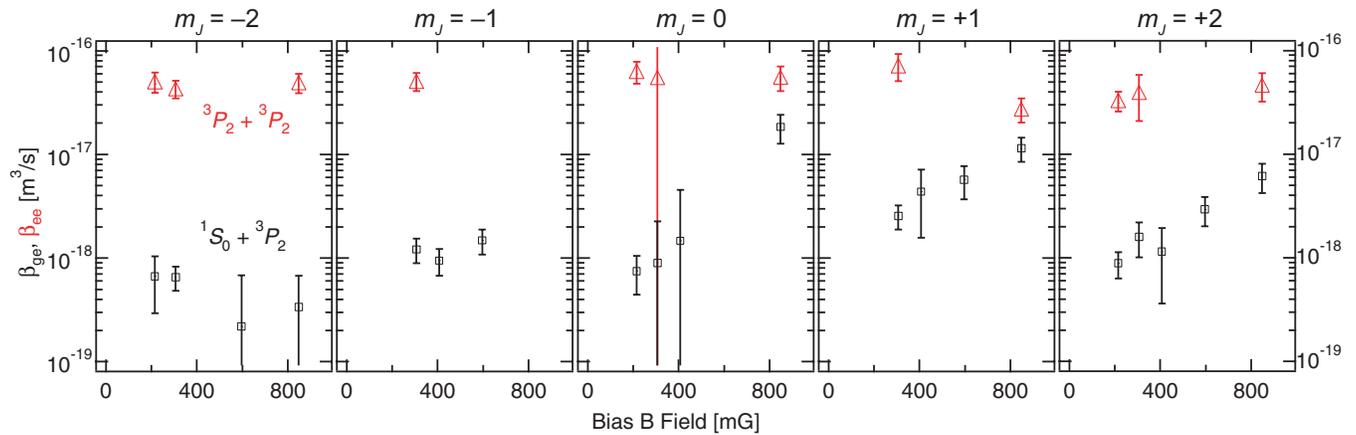}
 \caption{(color online) Inelastic decay rate for all $m_J$ states in
 various external magnetic fields. Red triangles show inelastic decay
 rate $\beta^{\text{in}}_{ee}$ of $^3P_2$--$^3P_2$ collision
 system. Essentially no spin-dependence is observed.
 Black squares show those of $^1S_0$--$^3P_2$ collision
 system ($\beta^{\text{in}}_{ge}$).
 Significant spin- and magnetic-field-dependence is observed.} 
\label{fig:rate}
\end{figure*}

The decay of $^3P_2$ state atoms in collisions dominantly with $^1S_0$
atoms is shown in Fig.~\ref{fig:decay}b. It is important to note that
in these decay curves the number of $^1S_0$ atoms is ten times the
number of $^3P_2$ atoms in either Fig.~\ref{fig:decay}a or
\ref{fig:decay}b. Thus, although the decay curves are steeper in this
raw data, as will be explained in detail later and shown in
Fig.~\ref{fig:rate}, $^1S_0$--$^3P_2$ collisions are less inelastic
than $^3P_2$--$^3P_2$ collisions. One can easily recognize in
Fig.~\ref{fig:decay}b that the $^1S_0$--$^3P_2$ collisional decay
behavior is markedly different from that of $^3P_2$--$^3P_2$
collisions.
The observed decay curves are strongly state-dependent.
In particular, the atoms in higher Zeeman levels in the $^3P_2$ state
($m_J \geq 0$) show a stronger field dependence in the decay. 
This clearly suggests the important role of $m$-changing collisions in
the $^1S_0$--$^3P_2$ collision. 

We extract 2-body inelastic loss rate coefficients from the observed
decay rates using the following method.
The decay is modeled by coupled differential equations:
\begin{equation}
\begin{split}
\dot n_e &= -\beta_{ge} n_g n_e - \beta_{ee} n_e^2, \\
\dot n_g &= -\beta_{ge} n_g n_e,
\end{split}
\label{eq:rate}
\end{equation}
where $n$ denotes the local density of atoms,
subscript $g$ ($e$) denotes ground (excited) state,
$\beta_{ge}$ ($\beta_{ee}$) is the two body 
loss coefficient for ground-excited (excited-excited) state collisions.
The trap lifetime for the ground state atoms is measured to be 30~s,
thus we neglect background collision loss because it is
much longer than the present two-body collision loss. 
For the $^3P_2$--$^3P_2$ collision, Eqs. \eqref{eq:rate} become simple and are solved analytically, since all atoms in the ground
state are removed from the trap($n_g=0$). 
By spatially integrating the equation, we obtain the evolution in atom
number 
\begin{equation}
N_e(t) = \frac{1}{1/N_{0e} +  G_{ee} t}.
\label{eq:Ne_3p2_3p2}
\end{equation}
Here $N_{0e}$ is the initial excited state atom number
and $G_{ee}=(V_{2e}/V_{1e}^2)\beta_{ee}$.
The effective volume 
is defined by
\begin{equation}
V_{qe} = \int d^3r [n_e(\mathbf{r})/n_{\text{0e}}]^q,
\end{equation}
where $q$ is an integer number, $n_{\text{0e}}$ is 
the peak density of excited state atoms,
and $n_e(\mathbf{r})$ is the spatial density distribution.

The spatial density distribution depends on 
the dimensionless parameter $\eta$: the ratio of the trap depth
$\epsilon_t$ to the sample temperature $k_B T$ in energy unit.
In our experiment, $\eta$ is 10 for the ground state atoms.
The trap depth for each $m_J$ component in $^3P_2$ state strongly
depends on the direction of external magnetic field and  
the polarization of FORT laser \cite{TIdo:PRL2003,AYamaguchi:NewJPhys2010}. 
To determine $\eta$ of the $^3P_2$ state,
we measure the Stark shift for all $m_J$ states 
as a function of horizontal FORT laser power
in various bias magnetic fields studied here.
The ratio $U_e/U_g$, where $U_{e(g)}$ is the trap potential of the
$^3P_2$ ($^1S_0$) state, 
varies from 1.02 to 1.31.
From the measurement, $\eta$ for all $m_J$ states
are calculated to be $7.2$ to $11.7$.
Since $\eta$ keeps more than 7 for any $m_J$ components,
large-$\eta$ approximation can be used to calculate 
the effective volume. 
In the large-$\eta$ limit, $n(\mathbf{r})$ is well approximated by the
thermal density distribution \cite{OJLuiten:PRA1996},
$
n_e(\mathbf{r}) = n_{0e} \exp[-U(\mathbf{r})/k_B T],
$
where $U(\mathbf{r})$ is the trap potential.
By approximating FORT potential to a truncated harmonic trap
$U(r)=\epsilon_t(r/R_0)^2 \Theta(R_0-r)$, the effective volume can be
written as  
\begin{equation}
V_q = R_0^3 \left(\frac{\pi}{q \eta}\right)^{3/2},
\end{equation}
where $\Theta(x)$ is the Heaviside step function, $U(R_0)=\epsilon_t$
is the trap depth, and $R_0$ is the boundary.

We can extract the $\beta_{ge}$ coefficient by analyzing the decay of
$^3P_2$ atoms in the presence of $^1S_0$ atoms, shown in
Fig.~\ref{fig:decay} (b).  
The data involves 
the $^1S_0$--$^3P_2$ collision as well as the
$^3P_2$--$^3P_2$. Thus, the analysis is not so simple as 
Eq. \eqref{eq:Ne_3p2_3p2}, because Eqs. \eqref{eq:rate} cannot in
general be solved analytically. 
However, if we assume depletion of the number of ground
state atoms is negligibly small, 
Eqs. \eqref{eq:rate} can be solved analytically.
In this case, we obtain the atom number evolution in a large-$\eta$
approximation 
\begin{equation}
N_e^{\text{A}}(t) = \frac{\exp(-G_{ge} N_{0g} t)}
 {\frac{1}{N_{0e}}+\frac{1}{N_{0g}}\frac{G_{ee}}{G_{ge}}
 [(1-\exp(-G_{ge} N_{0g} t)]},
\label{eq:anNe_1s0_3p2}
\end{equation}
where $G_{ge}=\beta_{ge}/\left(V^{2/3}_{1e}+V^{2/3}_{1g}\right)^{3/2}$
and $N_{0g}$ is the initial ground state atom number.
Note that we can reproduce Eq. \eqref{eq:Ne_3p2_3p2} from
Eq. \eqref{eq:anNe_1s0_3p2} if we set $N_{0g}=0$.  

We first fit the $^3P_2$--$^3P_2$ collision data by using
Eq. \eqref{eq:Ne_3p2_3p2} 
and calculate $\beta^{\text{in}}_{ee}$.
Then decay data of $^1S_0$--$^3P_2$ is fitted by using Eq. 
\eqref{eq:anNe_1s0_3p2} with fixed $\beta^{\text{in}}_{ee}$.
As mentioned in Ref. \cite{RdeCarvalho:PRA2004}, the observed two-body
decay rate $\beta$ includes inelastic collision loss
$\beta^{\text{in}}$ and evaporation $\beta^{\text{el}}$ due to elastic
collision; i.e., $\beta = \beta^{\text{in}} + f\beta^{\text{el}}$,
where $f$ represents the fraction of elastic collisions.
The inelastic collision rate  can be expressed as
$\beta^{\text{in}}=\beta/(f\gamma+1)$, where 
$\gamma$ is the ratio of elastic to inelastic cross sections, 
$\gamma \equiv \sigma^{\text{el}}/\sigma^{\text{in}}$
\cite{RdeCarvalho:PRA2004}.
In a large-$\eta$ approximation, 
$f$ can be approximated to a simple analytic 
equation \cite{MYan:arXiv2009}.
In the present condition, $\beta^{\text{in}}/\beta$ varies from
$0.87$ to $0.93$.

The inelastic collision rates for all $m_J$ states in various external
magnetic fields are plotted in Fig.~\ref{fig:rate}. 
Essentially no spin-dependence in the collision of two  metastable
$^3P_2$ states is apparent, which is represented by triangles in
Fig.~\ref{fig:rate}. 
In particular, the spin states of $m_J > -2$ show almost the same
inelastic collision rates with the lowest energy state of $m_J=-2$,
which should not suffer from the $m$-changing collision.
This behavior is quite different from the significant spin dependence
theoretically predicted in the collision of two metastable $^3P_2$
state at a high magnetic field 
\cite{VKokoouline:PRL2003,RSantra:PRA2003}.
Therefore, our result clearly shows that the $m$-changing
collision is not observed for all spin states at
this low magnetic field and low temperature. 
Since the most likely decay process for the $m_J=-2$ state is
the $J$-changing collision from the $^3P_2$ state to $^3P_0$ or
$^3P_1$, this would be dominant.
PQNC collision may also contribute to the overall decay. 
However it is difficult to distinguish $J$-changing from
PQNC collision unless we measure $\beta_{ee}$ for 
$^3P_0$--$^3P_0$ collision, a task beyond the scope of this Letter. 

Note that the average of the obtained inelastic collision rate
coefficients $\beta_{ee}$ of $4\times 10^{-17}$~m$^3/$s is 
a little higher than
the value of $1.0\times 10^{-17}$~m$^3/$s, which is previously
obtained at a 100 times higher temperature \cite{AYamaguchi:PRL2008}. 
This interesting temperature dependence is open to a further
theoretical investigation. 
In contrast, the significant spin- and field-dependence in the collision of the metastable $^3P_2$ state with the ground state $^1S_0$ is observed, which is represented by squares in Fig.~\ref{fig:rate}.
Especially, the spin states with higher energies ($m_J > -2$) show
higher inelastic collision rates compared with the lowest energy state
of $m_J=-2$.
Since the $m_J=-2$ state differs from other spin states only in that
the $m_J=-2$ state does not suffer from the $m$-changing collision, 
it is natural to think that the dominant decay process is a
$m$-changing collision in this $^3P_2$--$^1S_0$ collision at a low
field. 
The quantitative theoretical explanation of the detailed behaviors is
an interesting future work \cite{Tscherbul}.  

In addition, 
we can claim that the $m_J=-2$ state is rather stable
against the collision with the $^1S_0$ atom.
Since the decay of the $^3P_2$ ($m_J=-2$) atoms is dominated by the
$^3P_2$--$^3P_2$ collision even in the presence of the $^1S_0$ atoms,
as is shown in Fig. \ref{fig:rate}, it is quite difficult to
accurately extract the $\beta_{ge}$ coefficient for $m_J=-2$
state. Although our analysis results in the value of $\beta_{ge}$ on
the order of $10^{-19}$~m$^3/$s for the $m_J=-2$ state, it may be
much lower. In fact, our recent measurement on 
$^1S_0$--$^3P_2$ ($m_J=-2$) atoms in 
a tightly confined three dimensional optical lattice 
indicates that 
the upper limit of $\beta_{ge}$ is 
on the order of $10^{-20}$~m$^3/$s
\cite{SSugawa:inPreparation}.
This is promising to use in many applications. 

In conclusion, we have experimentally investigated collisional properties of spin polarized metastable $^3P_2$ states of Yb Atoms.
We reveal the significant spin-dependence in the collision of the metastable $^3P_2$ state with the ground state $^1S_0$, which strongly suggests
that the dominant decay process is a $m$-changing collision.
On the contrary, we observe essentially no spin-dependence in the
collision of two  metastable $^3P_2$ states, which is consistent with
the $J$-changing collision as a dominant decay process. 
Our results will trigger theoretical efforts to clarify these
behaviors quantitatively and give a guideline to use the  $^3P_2$
states in many possible applications.

The authors acknowledge 
very helpful experimental assistance of
 S. Sugawa, S. Kato, and 
Yb experiment team in Kyoto university.
We also thank T. Tscherbul, P. Zhang, and A. Dalgarno 
for helpful discussions.
This work was supported by the Grant-in-Aid for Scientific Research of
JSPS (No. 18204035, 21102005C01 (Quantum Cybernetics), 22684022), GCOE
Program ``The Next Generation of Physics, Spun from Universality and
Emergence'' from MEXT of Japan, and FIRST. 
One of us, J. M. D., thanks the Fulbright program for support.

\end{document}